\documentclass[]{iopart}

\begin{document}
\title{On the existence of cosmological event horizons}
\author
{Sourav Bhattacharya and Amitabha Lahiri}
\address{
S. N. Bose National Centre for Basic Sciences, \\
Block JD, Sector III, Salt Lake, Kolkata -700098, India.\\
}
\eads{\mailto{sbhatt@bose.res.in}, \mailto{amitabha@bose.res.in}}

%%%%%%%%%%%%%%%%%%%%%%%%%%%%%%%%%%%%%%%%%%%%%%%%%%%%%%%%%%%%%%%
\begin{abstract} We show that, for general static or axisymmetric
stationary spacetimes, a cosmological Killing horizon exists only if
$R_{ab}n^{a}n^{b}< 0$ for a hypersurface orthogonal timelike
$n^{a}$, at least over some portion of the region of interest of
the manifold.  This implies violation of the strong energy
condition by the matter fields, the simplest example of which is
a positive cosmological constant.
\end{abstract}



\medskip

It is generally accepted that a positive cosmological constant
implies the existence of a cosmological horizon, i.e. an outer
event horizon. If a positive cosmological constant $\Lambda$ is
added into the Einstein equation, we find de Sitter space in the
absence of matter for a spatially homogeneous and isotropic
universe.  This solution exhibits an outer Killing
horizon~\cite{Kastor:1992nn}.  If the spacetime is assumed to be
spherically symmetric and static, or axisymmetric and stationary,
the solution to the vacuum Einstein equations is Schwarzschild-de
Sitter or Kerr-de Sitter~ \cite{Carter:1968ks}. When the positive
$\Lambda$ and the other parameters of these solutions obey certain
conditions between them, one gets stationary black hole space-times
embedded within a cosmological event horizon.  What happens if
there is matter? A sufficiently low matter density should produce a
perturbation on the de Sitter black hole background. How does this
perturbation affect the global properties of the spacetime? In
particular, is there still an outer (cosmological) event horizon?
More generally, what is the criterion for the existence of a
cosmological event horizon? We were unable to find in the
literature anything resembling an existence proof, so we decided to
construct one. The motivation to look for horizons in spacetimes
with a positive cosmological constant comes from recent
observations that our universe is very likely endowed with
one~\cite{Riess:1998cb, Perlmutter:1998np}.

The goal of this paper is to find the general conditions for which
a stationary spacetime has an outer cosmological horizon. We
consider two types of spacetimes, one static, and the other
stationary and axisymmetric. An inner (black hole) event horizon is
not assumed, although one may be present. We assume that there is
no naked curvature singularity anywhere in our region of interest.
This implies that the invariants of the stress-energy tensor are
bounded everywhere in our region of interest, and that in the
absence of an inner horizon any closed surface can be continuously
shrunk to nothing. We also assume that the weak energy condition
(WEC) is satisfied by the stress-energy tensor.  We assume the
existence of a null outer horizon and find the condition that the
stress-energy tensor has to fulfill for the Einstein equations to
hold.  We find that the strong energy condition must be violated by
the stress-energy tensor, at least over some part of the spacelike
region inside the outer horizon. While a positive cosmological
constant does this, we also find conditions on the stress-energy
tensor due to ordinary matter so that $\Lambda>0$ implies an outer
horizon.

Let us then start with a spacetime which is static in some region.
In this region the spacetime is endowed with a timelike Killing
vector field $\xi^a$,
\begin{eqnarray}
\nabla_{a}\xi_b+\nabla_b \xi_{a}=0,
\label{s1}
\end{eqnarray}
with norm $\xi_{a}\xi^a=-\lambda^2$. Since the spacetime is static,
$\xi^a$ is orthogonal to a family of spacelike hypersurfaces
$\Sigma$, and the Frobenius condition is satisfied,
\begin{eqnarray}
\xi_{[a}\nabla_{b}\xi_{c]}=0.
\label{s2}
\end{eqnarray}
A horizon of this spacetime is defined as the null hypersurface on
which $\lambda^2=0$~\cite{Gibbons:1977mu}. 

Starting from the Killing identity
\begin{eqnarray}
\nabla_a\nabla^a \xi_b=-R_{ab}\xi^a\,,
\label{s3}
\end{eqnarray}
and contracting both sides of Eq. (\ref{s3}) by $\xi^b\,,$ we
obtain
\begin{eqnarray}
\nabla_a\nabla^a \lambda^2=2R_{ab}\xi^a\xi^b-
2\left(\nabla_a\xi_b\right)
\left(\nabla^a\xi^b\right).
\label{s4}
\end{eqnarray}
On the other hand, we can use Killing's equation~(\ref{s1}) and the 
Frobenius condition~(\ref{s2}) to get
\begin{eqnarray}
\nabla_a \xi_b=\frac{1}{\lambda}\left(\xi_b\nabla_a \lambda-
\xi_a \nabla_b \lambda \right),
\label{s5}
\end{eqnarray}
which we substitute into Eq.~(\ref{s4}) to obtain
\begin{eqnarray}
\nabla_a\nabla^a \lambda^2=2R_{ab}\xi^a\xi^b+
4\left(\nabla_a\lambda\right)
 \left(\nabla^a\lambda\right).
\label{s6}
\end{eqnarray}

In order to project Eq.~(\ref{s6}) onto $\Sigma$, we consider 
the usual projector or the induced metric on $\Sigma$
\begin{eqnarray}
h_{a}{}^{b}=\delta_{a}{}^{b}+\lambda^{-2}\xi_a\xi^b.
\label{s7}
\end{eqnarray}
Let us also write $D_a$ for the induced connection on $\Sigma$.
Then for any $p$-form $\Omega$ whose projection on $\Sigma$ is
$\omega$, and which satisfies $\pounds_\xi\Omega =
0$~\cite{Lahiri:1993vg},
\begin{eqnarray}
\lambda\nabla_a \Omega^{a\cdots} =  D_a(\lambda
\omega^{a\cdots})\,. 
\end{eqnarray}
Choosing the 1-form $\nabla_a\lambda^2$ for $\Omega_a$ in this
equation, and using Eq.~(\ref{s6}), we find
\begin{eqnarray}
  D_a\left(\lambda D^a \lambda^2\right) 
  =  2\lambda\left[R_{ab}\xi^a\xi^b+2 \left(D_a \lambda\right)
    \left(D^a \lambda\right)\right]\,.
\label{s14}
\end{eqnarray}
We will integrate this equation over the space-like hypersuface
with the horizon as boundary. For all known solutions with a
horizon defined by $\lambda^2=0,$ each term on both sides of this
equation is finite everywhere on $\Sigma,$ including on the
horizon(s). However, since we have unspecified energy-momentum on
$\Sigma,$ we will not presume that these terms remain finite on the
horizons. We will assume instead that the left hand side, for
example, does not diverge at the horizon faster than some inverse
power of $\lambda$. Then we multiply both sides of Eq. (\ref{s14})
by $\lambda^n$, for some appropriate $n>0$, and integrate over the
spacelike hypersurface $\Sigma$ to find
\begin{eqnarray}
\fl \oint_{\partial\Sigma}\lambda^{n+1} D_a \lambda^2d\gamma^{(2)a} =
2\int_{\Sigma}\left[\lambda^{n+1} R_{ab}\xi^a\xi^b+ 
\left(n+2\lambda\right)\lambda^n
\left(D_a \lambda\right)\left(D^a \lambda\right)\right].
\label{s15}
\end{eqnarray}
The surface integral is over the boundary of $\Sigma$, i.e. the
outer horizon whose existence we have assumed, and also over an
inner horizon if it exists. On the horizons, $\lambda=0$. So if we
choose $n$ to be sufficiently large, each of the invariant terms
appearing in the right hand side of Eq. (\ref{s15}) remains bounded
as $\lambda \to 0$.

Then the surface integral over the horizons vanishes, and we get
\begin{eqnarray}
\int_{\Sigma}
\left[\lambda^{n+1}
R_{ab}\xi^a\xi^b+ (n+2\lambda)\lambda^n
 \left(D_a\lambda\right)
 \left(D^a\lambda\right)\right]=0.
\label{s16}
\end{eqnarray}
Furthermore, since we have assumed that there is no naked curvature
singularity anywhere on $\Sigma$, in the absence of an inner black
hole horizon we may freely shrink the inner boundary to a
non-singular point, so that the corresponding integral vanishes
again. Thus Eq. (\ref{s16}) also holds for non-singular spacetimes
without a black hole.

The second term in Eq.~(\ref{s16}) with a positive $n$ is a
spacelike inner product and hence positive definite over $\Sigma$,
so we must have a negative contribution from the first term
$R_{ab}\xi^a\xi^b$. In other words, the outer horizon or the
cosmological horizon will exist only if
\begin{eqnarray}
R_{ab}\xi^a\xi^b<0,
\label{s17}
\end{eqnarray}
at least over some portion of $\Sigma$, so that the total integral
in Eq.~(\ref{s16}) vanishes. Using the Einstein equations
\begin{eqnarray}
R_{ab}-\frac{1}{2}R g_{ab}=T_{ab},
\label{s18}
\end{eqnarray}
we see that the condition~(\ref{s17}) implies that the strong
energy condition (SEC) is violated by the energy-momentum tensor
\begin{eqnarray}
\left(T_{ab}-\frac{1}{2}Tg_{ab}\right)\xi^a\xi^b<0,
\label{s19}
\end{eqnarray}
at least over some portion of $\Sigma$. We know that a positive
cosmological constant $\Lambda$, appearing on the right hand side
of the Einstein equations as $-\Lambda g_{ab}$, violates the SEC.
We now split the total stress-energy tensor $T_{ab}$ as
\begin{eqnarray}
T_{ab}=-\Lambda g_{ab}+T^{\rm{N}}_{ab},
\label{s20}
\end{eqnarray}
where the superscript `N' denotes `normal' matter fields satisfying
the SEC.  Then Eq.~(\ref{s16}) becomes
\begin{eqnarray}
  \int_{\Sigma}\lambda^n
  \left[\lambda X^{\rm{N}}+ (n+2\lambda)
(D_a \lambda)( D^a \lambda) -
    \Lambda\lambda^3\right]=0. 
\label{s21}
\end{eqnarray}
$X^{\rm{N}}$ is a positive definite contribution from the normal
matter satisfying SEC. So for the cosmological horizon to exist, we
must have
\begin{eqnarray}
\int_{\Sigma}\lambda^{n+1}
\left[X^{\rm{N}}-\Lambda\lambda^2\right]<0.
\label{s22}
\end{eqnarray}
In other words, the cosmological constant term (with $\Lambda >0$)
has to dominate the integral if there is to be an outer horizon.
It is interesting to note that the observed values of $\Lambda$ and
matter densities in the universe satisfy this requirement. So would
a universe with $\Lambda>0$ in which all normal matter is
restricted to a finite region in space. This has relevance in
discussions of the late time behavior of black holes formed by
collapse.

This result can be generalized to stationary axisymmetric
spacetimes, in general rotating, which satisfy some additional
constraints.  The basic scheme will be the same as before. For the
spacetime we assume two commuting Killing fields $(\xi^a,~\phi^a)$,
\begin{eqnarray}
\nabla_{(a}\xi_{b)} &=& 0 =
 \nabla_{(a}\phi_{b)} \,,\\
\left[\xi, \phi\right]^a &=& 0\,.
\label{2k}
\end{eqnarray}
$\xi^a$ is locally timelike with norm $-\lambda^2$, whereas
$\phi^a$ is a locally spacelike Killing field with closed orbits
and norm $f^2$. We also assume that the vectors orthogonal to
$\xi^a$ and $\phi^a$ span an integral submanifold. In other words,
local coordinates orthogonal to $\xi^a$ and $\phi^a$ can be
specified everywhere on the spacetime. This, and the last condition
above, are the additional constraints mentioned earlier. We note
that known stationary axisymmetric spacetimes obey these
restrictions.

For a rotating spacetime, $\xi^a$ is not orthogonal to $\phi^a$, so
in particular there is no spacelike hypersurface tangent to
$\phi^a$ and orthogonal to $\xi^a$. Let us first construct a family
of spacelike hypersurfaces. If we define $\chi_{a}$ as
\begin{eqnarray}
\chi_a=\xi_a-\frac{1}{f^2}\left(\xi_b\phi^b\right)
\phi_a \equiv \xi_a+\alpha \phi_a,
\label{sa2}
\end{eqnarray}
we will have $\chi_a\phi^a=0$ everywhere. An orthogonal basis for
the spacetime can be written as
$\left\{\chi^a,~\phi^a,~\mu^a,~\nu^a\right\}$. We note that
\begin{eqnarray}
\chi_a\chi^a=-\beta^2=
-\left(\lambda^2+\alpha^2 f^2\right),
\label{sa3}
\end{eqnarray}
i.e., $\chi_a$ is timelike when $\beta^2>0$. We can also calculate
that 
\begin{eqnarray}
\nabla_{(a}\chi_{b)} = \phi_a\nabla_b \alpha
+\phi_b\nabla_a \alpha.
\label{sa5}
\end{eqnarray}
Our assumption that $\{\mu^a,\, \nu^a\}$ span an integral 2-manifold
implies that
\begin{eqnarray}
\chi_{[a}\phi_b \nabla_c\phi_{d]} &=& 0\,,\\
\phi_{[a}\chi_b \nabla_c\chi_{d]} &=& 0\,.
\label{frobenius}
\end{eqnarray}
where we have also used Eq.~(\ref{sa2}). A straightforward
calculation from here shows that
\begin{eqnarray}
\nabla_a\chi_b-\nabla_b\chi_a=
2\beta^{-1}\left(\chi_b\nabla_a \beta-
\chi_a\nabla_b \beta\right).
\label{sa8}
\end{eqnarray}
It follows that $\chi^a$ satisfies the Frobenius condition,
\begin{eqnarray}
\chi_{[a}\nabla_b\chi_{c]}=0,
\label{sa4}
\end{eqnarray}
so there is a family of spacelike hypersurfaces $\Sigma$ orthogonal
to $\chi^a$, although we should note that $\chi^a$ is not a Killing
vector field. In a rotating black hole spacetime, $\xi^a$ becomes
spacelike within the ergosphere~\cite{Wald:1984rg}, so for such
spacetimes $\lambda^2=0$ does not in general define a horizon.  The
horizons are now located at $\beta^2=0$, as we will justify below.

Since $\chi^a$ is not a Killing vector --- $\alpha$ in
Eq.~(\ref{sa2}) is not a constant --- we need to ask if $\beta^2 =
0$ is a Killing horizon, i.e. if there is a Killing vector which
becomes null on the surface $\beta^2 = 0$. In order to understand
the nature of the surface $\beta^2 = 0$, let us consider the
congruence of null geodesics on this surface. We start by
constructing a null geodesic on the surface $\beta^2 = 0$.

The normal to a hypersurface defined by $u = 0$ is proportional to
$\nabla_a u$. Since the vector field $\chi^a$ is hypersurface
orthogonal as we have seen in Eq. (\ref{sa4}), we can
write~\cite{Gourgoulhon:2005ng}
\begin{eqnarray}
\chi_a=e^{\rho}\nabla_a u,
\label{k}
\end{eqnarray}
for some $\rho$ and $u$.

A straightforward computation using Eq. (\ref{sa5}) then shows
that, when $\beta^2 = 0$,  
\begin{eqnarray}
\chi^a\nabla_a \chi_b =
\frac12\nabla_b{\beta^2} 
= \kappa \chi_b,
\label{k1}
\end{eqnarray}
where $\kappa:=\pounds_{\chi}\rho$ is a function over that
surface. Eq. (\ref{k1}) shows that the 1-form $\nabla_a \beta^2$,
which is normal to the $\beta^2=0$ surface, is null on that
surface.  Eq. (\ref{k1}) also yields $\pounds_{\chi} \kappa=0$.

Now we can define a null geodesic $k^a$ $\left({\mathrm {i.e.}}\,,
  k_ak^a=0;~k^a\nabla_a k^b=0\right)$, tangent (or normal) to the
surface $\beta^2=0$, by $k^a=e^{-\kappa
  \tau}\chi^a$~\cite{Wald:1984rg}, where $\tau$ is the parameter
along $\chi^a$ satisfying $\chi^a\nabla_a \tau=1$. Then using
Eq.s (\ref{sa5}) and (\ref{sa4}) we find that $k^a$ satisfies
\begin{eqnarray}
\fl k_{[a}\nabla_{b]}k_c=e^{-2\kappa \tau}
\left[\frac{1}{2}\chi_{(a}\phi_{b}\nabla_{c)}\alpha-
\chi_c\nabla_a\chi_b-\chi_b\phi_a\nabla_c \alpha
-\chi_b\phi_c\nabla_a \alpha-
\chi_{c}\chi_{[a}\nabla_{b]}\left(\kappa \tau\right)
\right].\nonumber \\
\label{k2}
\end{eqnarray}
We next consider the Raychaudhuri equation for the null geodesic
congruence $\{k^a\}$,
\begin{eqnarray}
\frac{d\theta}{d\Theta}=-\frac{1}{2}\theta^2
-\hat{\sigma}_{ab}\hat{\sigma}^{ab}+\hat{\omega}_{ab}
\hat{\omega}^{ab}-R_{ab}k^ak^b,
\label{k3}
\end{eqnarray}
where $\Theta$ is the parameter along the geodesic $k^a$; $\theta$,
$\hat{\sigma}_{ab}$ and $\hat{\omega}_{ab}$ are respectively the
expansion, shear and rotation of the congruence defined by
\begin{eqnarray}
\theta=\hat{h}^{ab}\widehat{\nabla_a k_b};\qquad
\hat{\sigma}_{ab}=\widehat{\nabla_{(a} k_{b)}}-\frac{1}{2}\theta
\hat{h}_{ab};\qquad
\hat{\omega}_{ab}=\widehat{\nabla_{[a} k_{b]}}.
\label{expressions}
\end{eqnarray}
The `hat' over the tensors denotes that they are evaluated on a
spacelike 2-plane orthogonal to $k^a$ (or $\nabla_a \beta^2$), and
$\hat{h}_{ab}$ is the metric on this plane (see e.g.
\cite{Wald:1984rg, Hawking:1973uf} for details on null congruence).
Tangent to the $\beta^2=0$ surface (i.e., orthogonal to $\nabla_a
\beta^2$), we can choose $\phi^a$ as a basis vector on the
spacelike 2-plane. Let the other basis vector be some $X^a$, with
$\phi_a X^a=0$, and of course $\phi_a k^a = 0 = X_a k^a$. Also,
since $\phi^a$ is a Killing field and commutes with $\xi^a$, we
have $\pounds_{\phi}\alpha=0$.

We now contract Eq. (\ref{k2}) by
$(\phi^b\phi^c+X^bX^c)$ to find that 
\begin{equation}
k_a(\phi^c\phi^b+X^cX^b)\nabla_b k_c=0\,,
%\label{}
\end{equation}
which implies the expansion $\theta=0$ on the surface $\beta^2=0$,
and thus the left hand side of Eq.~(\ref{k3}) is also zero.
Similarly by contracting Eq.~(\ref{k2}) by $\phi^{[b}X^{c]}$, we
see that the rotation $\hat{\omega}_{ab}$ also vanishes. However if
we contract Eq. (\ref{k2}) by $\phi^{(b}X^{c)}$, we see that the
components of the shear $\hat{\sigma}_{ab}$ do not vanish,
\begin{eqnarray}
k_a \phi^{(b}X^{c)}\hat\sigma_{bc}=
\frac{1}{2}e^{-\kappa \tau}\phi^{(b}X^{c)}
\phi_c\left(\nabla_b \alpha\right) k_a,
\label{shear}
\end{eqnarray}
where we have used the fact that $\theta=0$. Since the Ricci scalar
$R$ is finite at $\beta^2=0$ by assumption, Eq.~(\ref{k3}) becomes
upon using the Einstein equations
\begin{eqnarray}
T_{ab}k^ak^b=-\frac14e^{-2\kappa \tau}
f^2\left(\widehat{\nabla}_a\alpha\right)
\left(\widehat{\nabla}^a\alpha\right).
\label{k4}
\end{eqnarray}
The inner product on the right hand side of Eq. (\ref{k4}) is
spacelike. So we see that the null energy condition could be
violated on $\beta^2=0$,
\begin{equation}
T_{ab}\chi^a\chi^b\leq0.
%\label{}
\end{equation}
This also implies by continuity the violation of the WEC close to
the hypersurface $\beta^2=0$. Since by our assumption we are not
violating WEC, we must have $\left(\widehat{\nabla}_a\alpha\right)
\left(\widehat{\nabla}^a\alpha\right) =0$ on the spacelike section
of the $\beta^2=0$ surface. On the other hand, using Eq.~(\ref{2k})
we see that $\pounds_{\chi}\alpha=0$ everywhere. Thus $\alpha$ is a
constant on the $\beta^2=0$ surface. Therefore, on this surface,
$\chi^a$ coincides with a Killing vector field, and hence the
horizons we have defined are Killing horizons.  This is actually an
old result~\cite{Carter:1969zz}, which we have rederived using an
alternative method.

After this necessary digression, we return to our main proof of
existence of horizons.  Using the Killing identities
$\nabla_{a}\nabla^a\xi_b= -R_{b}{}^{a}\xi_a$, and
$\nabla_{a}\nabla^a\phi_b=-R_{b}{}^{a}\phi_a$, and also the
orthogonality $\chi_a\phi^a=0$, we obtain on $\Sigma$
\begin{eqnarray}
\chi^b\nabla_a\nabla^a\chi_b
=-R_{ab}\chi^a\chi^b
+2\chi^a\nabla_c\phi_a \nabla^c \alpha \,,
\label{sa6}
\end{eqnarray}
which is equivalent to
\begin{eqnarray}
\nabla_a\nabla^a\beta^2
=2R_{ab}\chi^a\chi^b-2\nabla^c\chi^a\nabla_c\chi_a
-4\chi^a\nabla_c\phi_a\nabla^c \alpha.
\label{sa7}
\end{eqnarray}
Note that if we set $\alpha=0$ in Eq.~(\ref{sa7}), we recover
the static case of Eq.~(\ref{s4}).  

Next we note that the subspace spanned by
$\left\{\chi^a,~\mu^a,~\nu^a \right\}$ do not form a hypersurface.
This is because the necessary and sufficient condition that an
arbitrary subspace of a manifold forms an integral submanifold or a
hypersurface is the existence of a Lie algebra of the basis vectors
of that subspace (see e.g.~\cite{Wald:1984rg} and references
therein).  The condition Eq.~(\ref{sa4}) follows from this. On the
other hand, Lie brackets among $\left\{\chi^a,~\mu^a,~\nu^a
\right\}$ do not close. For example,
\begin{eqnarray}
[\chi,~\mu]^a=[\xi,~\mu]^a+\alpha[\phi,~\mu]^a
+\phi^a \mu^b\nabla_b \alpha.
\label{sa10}
\end{eqnarray}
Since $\mu^a$ is not a Killing field, the last term on the right
hand side of Eq.~(\ref{sa10}) is not zero. A similar argument holds
for $\nu^a\,.$ Therefore the vectors spanned by
$\left\{\chi^a,~\mu^a,~\nu^a\right\}$ do not form a Lie
algebra. This implies that we cannot write a condition like
$\phi_{[a}\nabla_b\phi_{c]}=0$.

However, according to our assumptions, there are integral spacelike
$2$-manifolds orthogonal to both $\chi^a$ and $\phi^a$. These are
spanned by $\left\{\mu^a,\, \nu^a\right\}$. Then we must have
\begin{eqnarray}
\phi_{[a}D_{b}\phi_{c]}=0,
\label{sa11}
\end{eqnarray}
where $D_b$ is the connection induced on $\Sigma$ defined via the
projector $h_{a}{}^{b}=\delta_{a}{}^{b}+\beta^{-2}\chi_a\chi^b$,
exactly in the same manner as in the static case. Then we can write
\begin{eqnarray}
D_a\phi_b=h_{a}{}^{c}h_{b}{}^{d}
\nabla_c\phi_d
=\nabla_a \phi_b +\beta^{-2}
\left(\chi_a \phi^c\nabla_b \chi_c
-\chi_b \phi^c\nabla_a \chi_c \right).
\label{sa12}
\end{eqnarray}
Using the expression of $\nabla_a\chi_b$ from Eq.~(\ref{sa5}) and
Eq.~(\ref{sa8}), we can rewrite this as
\begin{eqnarray}
D_a\phi_b=\nabla_a\phi_b+\frac{f^2}{2\beta^2}
\left[\chi_a \nabla_b \alpha
-\chi_b \nabla_a \alpha \right].
\label{sa13}
\end{eqnarray}
It follows from this equation that we can write the Killing
equation for $\phi_a$ on $\Sigma$ as
\begin{eqnarray}
D_a\phi_b + D_b\phi_a=0.
\label{sa14}
\end{eqnarray}
Using this equation and the Frobenius condition of
Eq.~(\ref{sa11}), we derive the expression
\begin{eqnarray}
\nabla_a\phi_b=\frac{1}{f}
\left[\phi_b D_a f-\phi_a D_b f\right]+
\frac{f^2}{2\beta^2}
\left[\chi_b \nabla_a \alpha
-\chi_a \nabla_b \alpha \right].
\label{sa16}
\end{eqnarray}
These are all that is needed to simplify Eq. (\ref{sa7}).
Substituting the expressions for $\nabla_a \chi_b$ and $\nabla_a
\phi_b$ into Eq. (\ref{sa7}) we get
\begin{eqnarray}
\nabla_a\nabla^a \beta^2=2R_{ab}\chi^a\chi^b+
4\left(\nabla_a\beta\right)
\left(\nabla^a\beta\right)
+f^2\left(\nabla_a 
\alpha\right)\left(\nabla^a \alpha\right).
\label{sa17}
\end{eqnarray}
With this, using the same line of argument as for Eq.~(\ref{s21}),
we get
\begin{eqnarray}
\fl
\int_{\Sigma}\beta^n
\left[\beta X^{\rm{N}}+ (n+2\beta)
\left(D_a \beta\right)
\left( D^a \beta\right)
+\frac{f^2\beta}{2}\left(D_a \alpha\right)
\left(D^a \alpha\right) 
 -\Lambda\beta^3\right]=0,
\label{sa18}
\end{eqnarray}
if the spacetime has an outer or cosmological event horizon. For
$T_{ab}=0$ in Eq. (\ref{sa18}), we get Kerr-de Sitter solution
\cite{Carter:1968ks}. We note that the assumption of integral
2-manifolds orthogonal to both the Killing fields $\xi^a$ and
$\phi^a$ was crucial to the proof. For a completely general
stationary axisymmetric spacetime, the existence of such
submanifolds is not guaranteed, and thus an outer horizon may not
exist in such cases, even for $\Lambda >0$.

To summarize, we have found that in both the static and the
stationary axisymmetric cases, existence of an outer Killing
horizon requires a violation of the strong energy condition. This
can be through a positive cosmological constant, for which there is
strong observational evidence, or through exotic matter.

%%%%%%%%%%%%%%%%%%%%%%%%%%%%%%%%%%%%%%%%%%%%%%%%%%%%%%%%%%%
\section*{Acknowledgement}
The authors wish to thank the anonymous referees for insightful
questions and comments. S.~B. also wishes to thank A.~Ghosh for
useful discussions. S.~B.'s work was supported by a fellowship from
his institution SNBNCBS.

%%%%%%%%%%%%%%%%%%%%%%%%%%%%%%%%%%%%%%%%%%%%%%%%%%%%%%%%%
\vskip 1cm
%%%%%%%%%%%%%%%%%%%%%%%%%%%%%%%%%%%%%%%%%%%%%%%%%%%%%%%%%%%%%%%
%                 BIBLIOGRAPHY
%%%%%%%%%%%%%%%%%%%%%%%%%%%%%%%%%%%%%%%%%%%%%%%%%%%%%%%%%%%%%%%%%%%%

\end{document}